\documentclass[3p,times,procedia]{elsarticle}
\usepackage{nupha_ecrc}

\volume{00}

\firstpage{1}

\journalname{Nuclear Physics A}

\runauth{S.~Takeuchi et al.
}

\jid{nupha}

\jnltitlelogo{Nuclear Physics A}

\usepackage{amssymb}
\usepackage[figuresright]{rotating}

\usepackage{color}
\begin{document}

\begin{frontmatter}



\dochead{}
%% e.g. \dochead{17th International Conference on Dynamical Processes in Excited States of Solids}

\title{Violation of mass ordering for multi-strange hadrons at RHIC and LHC}

\author[label1]{Shiori Takeuchi}
\author[label1,label2,label3]{Koichi Murase}
\author[label1]{Tetsufumi Hirano}
\author[label4,label5,label6]{Pasi Huovinen}
\author[label7]{Yasushi Nara}

\address[label1]{Department of Physics, Sophia University, Tokyo 102-8554, Japan}
\address[label2]{Department of Physics, The University of Tokyo, Tokyo 113-0033, Japan}
\address[label3]{Theoretical Research Division, Nishina Center, RIKEN, Wako 351-0198, Japan}
\address[label4]{Institut f{\"u}r Theoretische Physik, Johann Wolfgang Goethe-Universit{\"a}t, 60438 Frankfurt am Main, Germany}
\address[label5]{Frankfurt Institute for Advanced Studies, 60438 Frankfurt am Main, Germany}
\address[label6]{Institute of Theoretical Physics, University of Wroclaw, 50-204 Wroclaw, Poland}
\address[label7]{Department of International Liberal Arts, Akita International University, Yuwa, Akita-city 010-1292, Japan}

\begin{abstract}
We study effects of the hadronic rescattering on final observables 
especially for multi-strange hadrons such as 
$\phi$, $\Xi$ and $\Omega$
in high-energy heavy-ion collisions 
within an integrated dynamical approach.
In this approach, (3+1)-dimensional ideal hydrodynamics is combined 
with a microscopic transport model, JAM. 
We simulate the collisions with or without hadronic rescatterings 
and compare observables between these two options
so that we quantify the effects of the hadronic rescattering.
We find that the mean transverse momentum and the elliptic flow parameter
of multi-strange hadrons are less affected 
by hadronic rescattering and, as a result, 
the mass ordering of the $p_T$-differential elliptic flow parameter $v_2(p_T)$ is violated: 
At the RHIC and the LHC energies the $v_2(p_T)$ for $\phi$-mesons is larger than that for protons
in the low-$p_T$ regions.
\end{abstract}


\end{frontmatter}

% \linenumbers

\section{Introduction}
\label{Introduction}
The main purpose in the physics of high-energy heavy-ion collisions 
at the Relativistic Heavy Ion Collider (RHIC) and the Large Hadron Collider (LHC) is 
to extract properties of the quark gluon plasma (QGP), 
the deconfined nuclear matter consisting of strongly interacting quarks and gluons. 
In particular, transport properties of nearly perfect QGP fluids attract a great deal of attention.

The QGP created in the collisions expands, cools down and 
finally turns into a hadron gas. 
Hadrons rescatter with each other in this late stage of the collision, 
thus information about the QGP is usually contaminated by
 the hadronic rescatterings.
This fact makes it difficult to observe the QGP directly.
For this reason it is suggested that multi-strange hadrons can be utilised 
as direct probes of the QGP.
Since the multi-strange hadrons have small cross sections with pions, 
the dominant constituents of a hadron gas, 
their distributions reflect the state of the system
just after hadronization~\cite{Shor:1984ui,vanHecke:1998yu,Bass:2000ib,Cheng:2003as,Hirano:2007ei,He:2011zx,Nasim:2012gz,Zhu:2015dfa}.
Unlike conventional penetrating probes such as photons and dileptons 
which are emitted during the entire evolution of the system,
the multi-strange hadrons provide
information about this specific stage of the collisions.

Hydro + cascade calculations predicted
several years ago that
the elliptic flow coefficient $v_2(p_T)$ for protons and $\phi$-mesons 
violates the mass ordering of this coefficient.
This phenomenon reflects the small scattering cross section of $\phi$-meson, 
and was recently observed by the STAR collaboration~\cite{Nasim:2012gz}.

In this contribution,
we study the violation of mass ordering more systematically and quantitatively 
by focusing on $p_T$ distributions and elliptic flow of hadrons, 
in particular for $\Xi$- and $\Omega$-baryons and $\phi$-mesons.
An integrated dynamical model, 
a more sophisticated version of the hydro + cascade approach, is employed here 
to make the investigation more realistic.

\section{Model}
\label{Model}
We simulate Au + Au collisions at $\sqrt{s_{\rm NN}}=200$ GeV 
and Pb + Pb collisions at $\sqrt{s_{\rm NN}}=2.76$  TeV
on an event by event basis 
by employing an integrated dynamical approach~\cite{Hirano:2012kj}.
This approach consists of three stages.
In the initial stage, entropy-density distribution after the collision is calculated 
by using a Monte Carlo Glauber model.
The subsequent QGP fluid expansion is described by fully (3+1) dimensional ideal hydrodynamics.
After we switch the description from fluids to particles, 
we utilise a hadron cascade model, JAM~\cite{Nara:1999dz}, 
to describe the evolution of hadron gas.
As for an equation of state (EOS),
we employ \textit{s95p}-v1.1~\cite{Huovinen:2009yb},
in which the lattice EOS at high temperature is connected 
to the hadron resonance gas EOS at low temperature.
Note that this particular version of the model EOS is designed to include
all the hadrons in JAM. 
Switching from hydrodynamics to JAM is done 
by using the Cooper-Frye formula
on the isothermal hypersurface at the temperature of 155 MeV.
This temperature is chosen 
to reproduce the experimentally observed pion-to-kaon and pion-to-proton ratios
in low $p_T$ regions at the RHIC energy.

Hadronic reactions in JAM are described as two-particle scatterings with 
experimental hadronic cross section if available.
When there are no experimental data, we employ the additive quark model
for the corresponding scatterings.
In this model, reactions involving (hidden-)strange hadrons have smaller cross sections
than non-strange hadrons due to a phenomenological strangeness suppression factor.
Furthermore, the experimentally known scattering cross sections of multi-strange hadrons are small,
since they form very few (or not at all) resonances.
Thus multi-strange hadrons have smaller cross sections than non-strange hadrons.
Note here that, 
in order to study the effects of hadronic rescattering on $\phi$-meson efficiently, 
we switch off the decay channel $\phi \rightarrow K^+ K^-$.
This does not affect the kinetic evolution of the system 
because the lifetime of $\phi$-mesons ($\sim$ 47 fm/$c$)~\cite{Agashe:2014kda} is larger than typical lifetime of the system ($\sim$ 10 fm/$c$).
For further details, see Ref.~\cite{Hirano:2012kj}.

\section{Results}
\label{Results}
To investigate the effects of hadronic rescattering on final observables, 
we simulate the collisions with two options in JAM. 
One of them is the default setting,
in which the rescatterings occur until all hadrons have decoupled,
and resonances decay according to their lifetimes and decay channels.
By using this option, we are able to reasonably reproduce final experimental
observables such as 
the $p_T$-spectra and differential $v_2$
at both the RHIC~\cite{Takeuchi:2015ana} and the LHC energies. 
In the other option, the hadronic rescatterings are deactivated but
resonances decay. 
These calculations serve the information just after the fluid-dynamical stage.
Comparisons of observables 
calculated using these two options show
how much the hadronic rescattering affects final observables.

The phenomenon of violation of mass ordering in $v_2(p_T)$ 
can be interpreted 
as a result of interplay between hadronic rescattering effects 
on mean transverse momentum, $\langle p_T \rangle$, and those on
 $p_T$-averaged $v_2$
because the slope of $v_2(p_T)$ is roughly approximated 
by the relation~\cite{Hirano:2005wx}, 
${dv_{2}(p_T)}/{ dp_T} \approx {v_{2}}/{\langle p_T \rangle}$.
Therefore we quantify the effects 
on $\langle p_T \rangle$ and $v_2$ for each hadron
by taking ratio of the observables just after the fluid stage to the final observables.
As shown in Fig.~8 (a) in Ref.~\cite{Takeuchi:2015ana}, 
the ratio of $\langle p_T \rangle$ for pions, kaons and protons 
follow the tendency obtained from $m_T$ scaling ansatz,
in which it is assumed that all the hadrons flow with common velocity.
However multi-strange hadrons obviously deviate from this pattern. 
From this observation, 
multi-strange hadrons do not fully participate in the radial flow during the hadronic stage
and therefore freeze out earlier than non-strange hadrons.
As for $v_2$ shown in Fig.~8 (b) in Ref.~\cite{Takeuchi:2015ana}, 
pion $v_2$ increases by about 20\% during the hadronic stage,
whereas the $v_2$ of all the other hadrons shows much smaller increase of 0-5\% .
By combining the results of these two observables, we see that 
both $\langle p_T \rangle$ and $v_2$ for multi-strange hadrons are hardly affected
by hadronic rescatterings, 
but either one of these two observables is affected for all the other particles.
This fact is reflected in $v_2(p_T)$ for each hadron.
In Fig.~6 in Ref.~\cite{Takeuchi:2015ana}, 
we showed $v_2(p_T)$ with or without hadronic rescatterings 
to see 
how rescatterings affect it.
$\phi$-meson $v_2(p_T)$
is almost identical in both cases 
since the hadronic rescatterings do not change its slope. 
However the situation is different in the case for non-strange hadrons.
Pion $v_2(p_T)$ goes up 
because $p_T$-averaged $v_2$ increases 
but $\langle p_T \rangle$ remains almost unchanged
in the hadronic stage.
On the other hand, for protons,
$p_T$-averaged $v_2$ does not change a lot, 
but  $\langle p_T \rangle$ increases.
Consequently, 
proton $v_2(p_T)$ shifts to higher $p_T$ region
and crosses $\phi$-meson 
$v_2(p_T)$ at $\sim1.5$ GeV
violating the conventional mass ordering.

\begin{figure*}[tbp]
\begin{center}
\includegraphics[bb=50 50 554 554,angle=-90,width=0.45\textwidth]{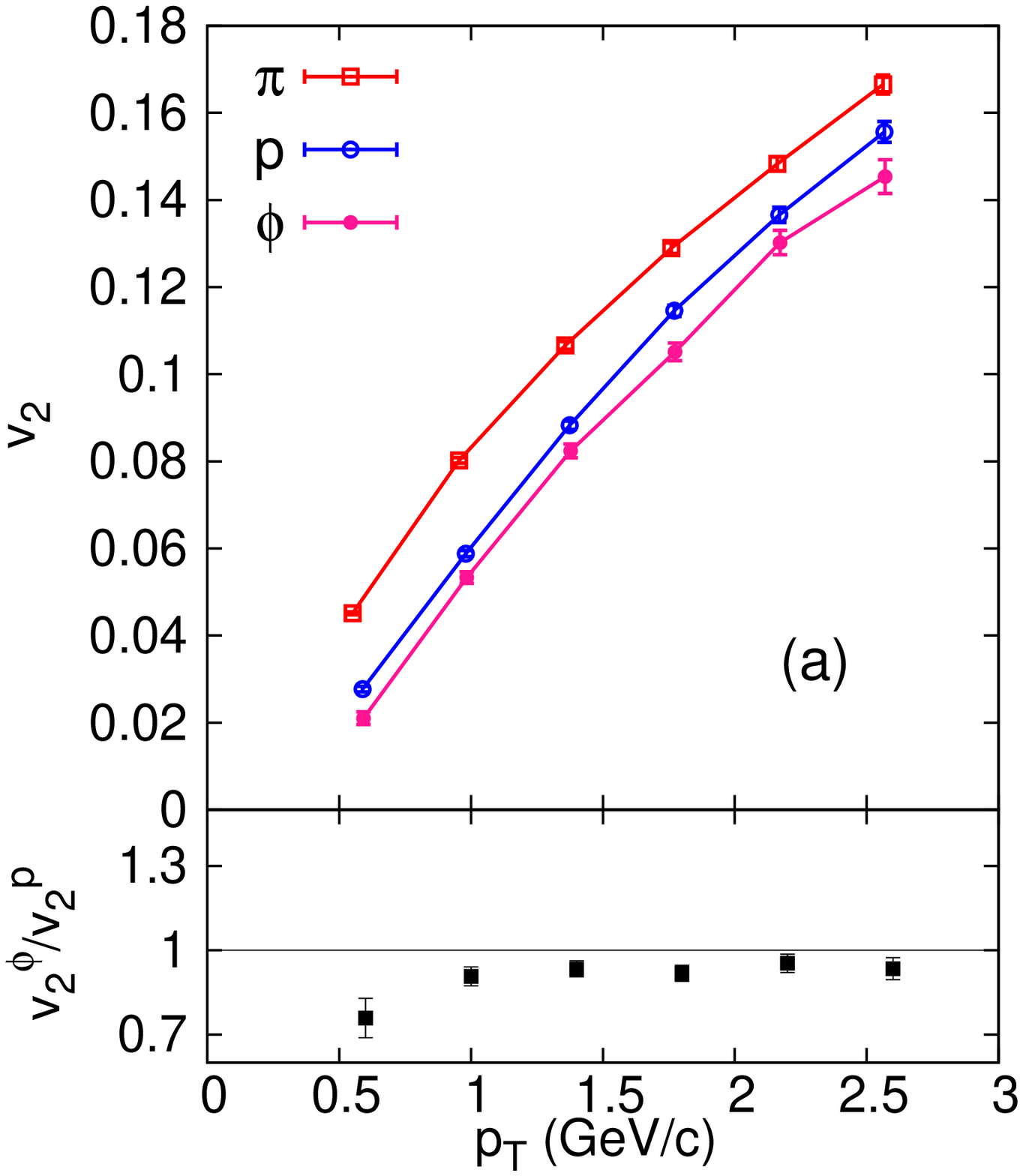}
\includegraphics[bb=50 50 554 554,angle=-90,width=0.45\textwidth]{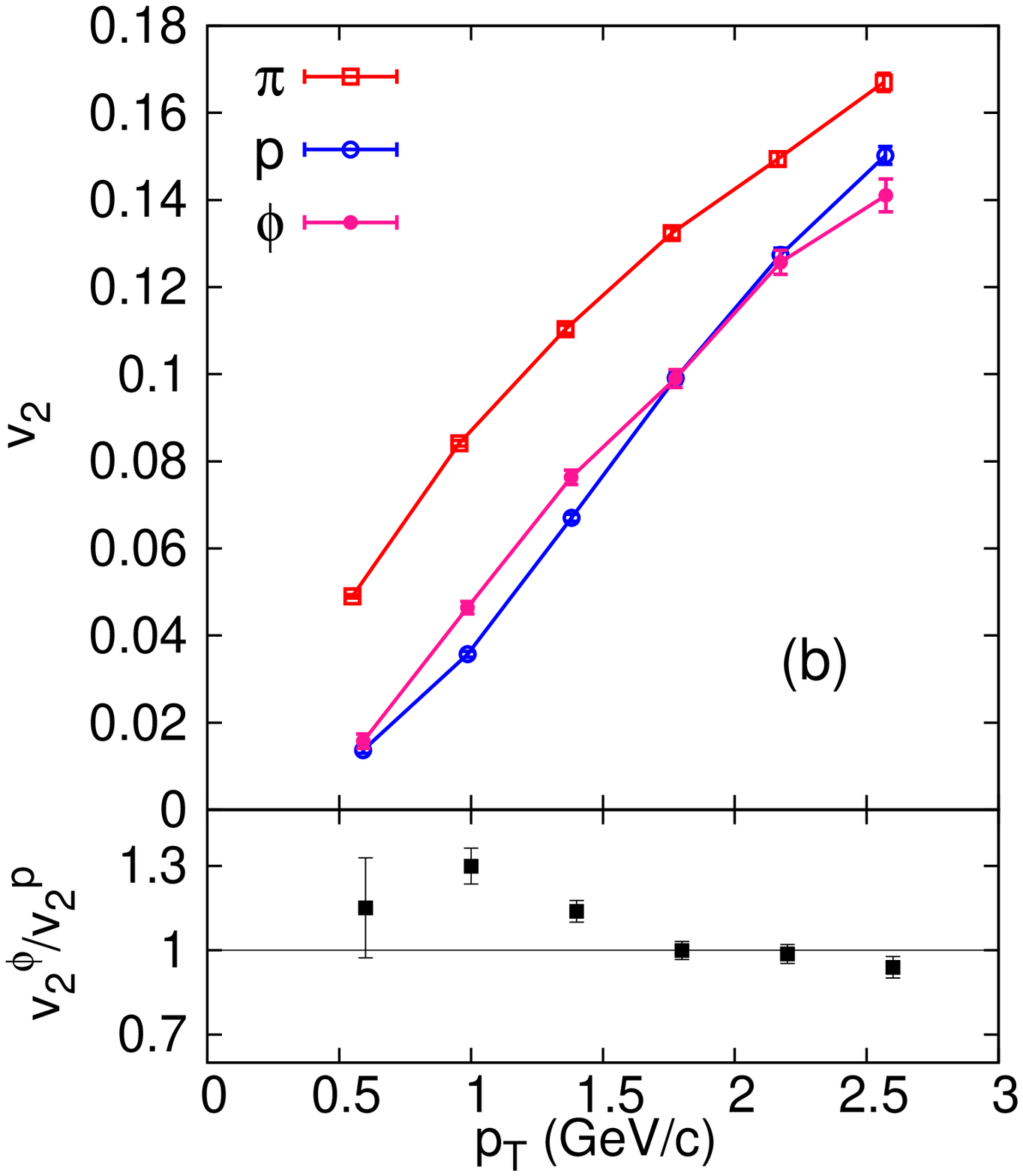}
\end{center}
\caption{Transverse momentum dependencies of
  elliptic flow parameter $v_2$ for pions (open square),  protons (open circle) 
  and $\phi$-mesons (filled circle) near midrapidity $\left| \eta \right| < 2.0$ obtained 
  from the integrated dynamical approach 
  (a) without hadronic rescattering and (b) with hadronic rescattering 
  in minimum bias  Pb+Pb collisions at $\sqrt{s_{\rm NN}}=2.76$ TeV. 
  The lower panels of the plots show the ratio of $v_{2}^{\phi}$ to $v_{2}^{p}$.}
\label{fig:v2_noscat_scat}
\end{figure*} 

At the LHC energy, this phenomenon appears in the same way.
Figure~\ref{fig:v2_noscat_scat} shows $v_2(p_T)$ for pions, protons and $\phi$-mesons
in minimum bias Pb+Pb collisions at $\sqrt{s_{\rm NN}}=2.76$ TeV 
from the integrated dynamical approach.
To see this behaviour clearly,
we also plot the ratio $v_2^{\phi}/v_2^{p}$
in lower panels of the figures.
In the case without hadronic rescatterings shown in Fig.~\ref{fig:v2_noscat_scat} (a),
the mass ordering behaviour,
$v_2^{\pi}(p_T) > v_2^{p}(p_T) > v_2^{\phi}(p_T)$
for $m_{\pi} < m_{p} < m_{\phi}$,
appears due to the collective flow in the fluid stage.
However, in Fig.~\ref{fig:v2_noscat_scat} (b),
this pattern is reversed below about 2 GeV between protons and $\phi$-mesons:
$v_2^{p}(p_T) < v_2^{\phi}(p_T)$ even though $m_{p} < m_{\phi}$.
These are qualitatively the same results to those at the RHIC energy
but quantitatively the crossing point between these two
shown in Fig.~\ref{fig:v2_noscat_scat} (b)  shifts 
to higher $p_T$ region 
compared to that at the RHIC energy as shown 
in Fig.~6 (b) in Ref.~\cite{Takeuchi:2015ana}.

\begin{figure*}[tbp]
\begin{center}
\includegraphics[clip,angle=-90,width=0.42\textwidth]{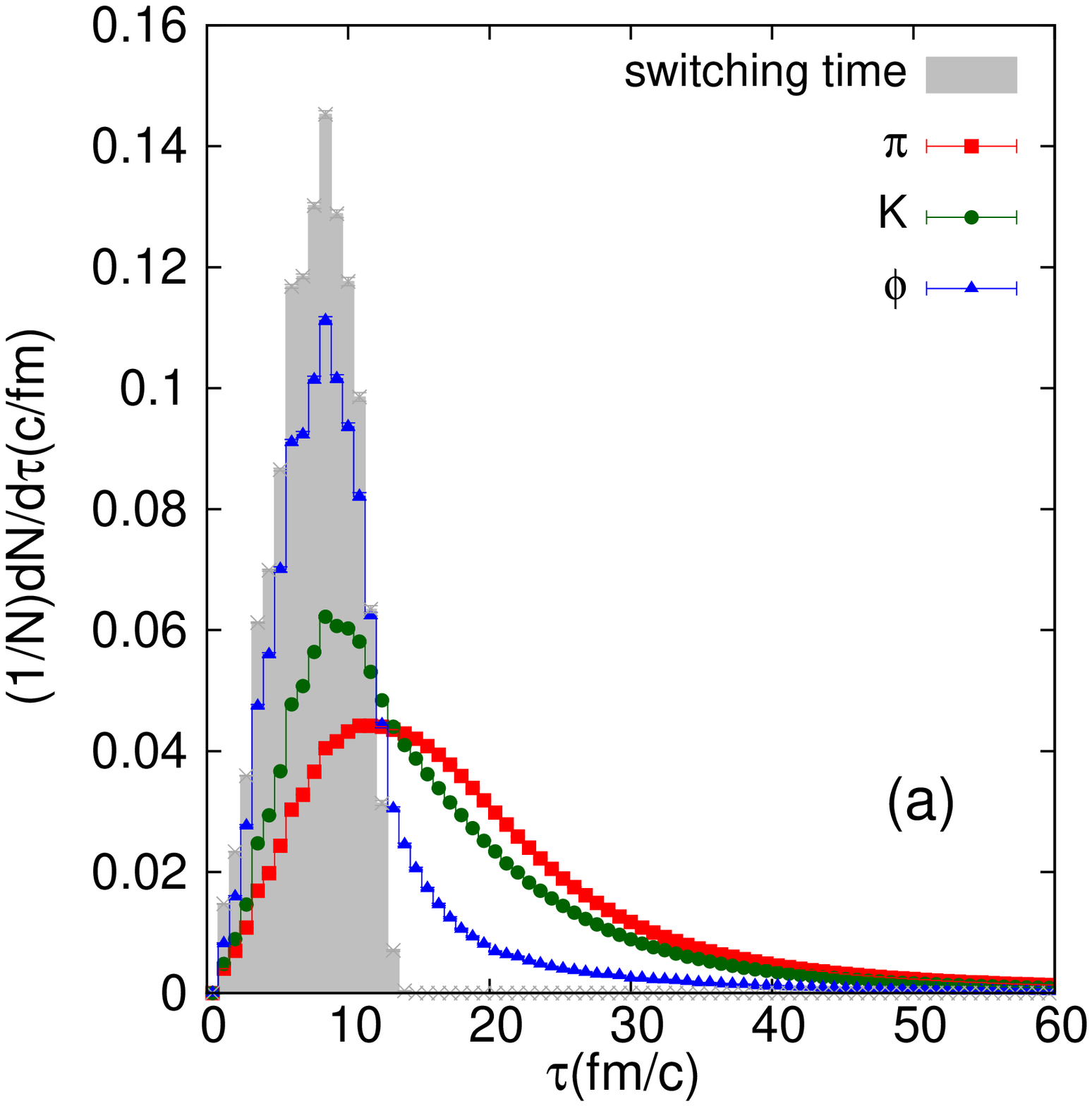}
\includegraphics[clip,angle=-90,width=0.42\textwidth]{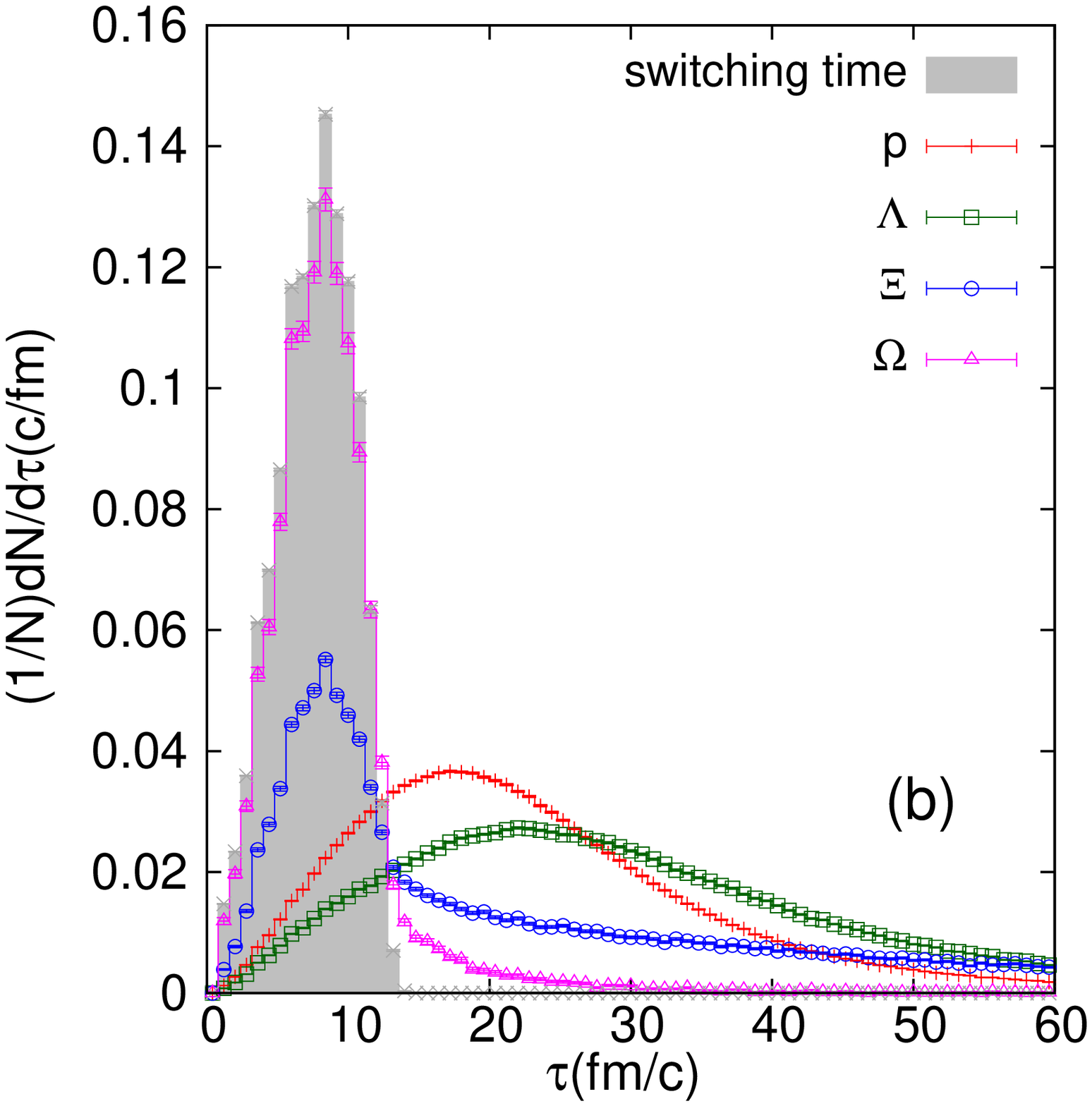}
\end{center}
\caption{Normalised 
 freeze-out time ($\tau$) distributions
 for (a) mesons ($\pi$, $K$ and $\phi$)
 and (b) baryons ($p$, $\Lambda$, $\Xi$ and $\Omega$)
 near midrapidity $\left| y \right| < 1.0$ in minimum bias
 Pb+Pb collisions at  $\sqrt{s_{\rm NN}}=2.76$ TeV.
The shaded areas represent
the distributions of charged hadrons 
at the time switching from fluids to particles.}
\label{fig:dNdtau}
\end{figure*}

In addition to these results, 
we also show the normalised freeze-out time distributions for identified hadrons. 
In Figs.~\ref{fig:dNdtau} (a) and (b),
we show the results for mesons and for baryons 
in separate panels for clarity.
Also switching time distributions from fluids to particles are shown with shaded areas.
Prominent peaks around 10 fm/$c$ 
for $\phi$-mesons and $\Omega$-baryons can be seen
and look quite similar to the ones of the switching time distributions.
The distribution for $\Xi$-baryons has also a peak in the early time but its height is lower
than for $\phi$-mesons and $\Omega$-baryons.
This is because the decay contribution from long-lived resonance 
$\Xi$(1530) to $\Xi$ forms a long tail in the late time.
Therefore primordial $\Xi$-baryons freeze out 
as early as $\phi$-mesons and $\Omega$-baryons.
These results prove that the multi-strange hadrons freeze out 
soon after the fluid stage
since they rarely rescatter in the hadronic stage.

\section{Summary}
\label{Summary}
We have studied the effects of the hadronic rescattering on
observables especially for multi-strange hadrons.
We have used an integrated dynamical approach, 
a model combining the ideal hydrodynamics with a hadronic cascade model, JAM.
In order to investigate the hadronic rescattering effects within this approach, 
we have compared 
$p_T$-distributions and elliptic flow
with hadronic rescatterings to the ones without rescatterings.
By studying the effects on mean transverse momentum 
and integrated elliptic flow parameters,
we have found that these observables for the multi-strange hadrons
 are less affected by the rescatterings.
Furthermore theoretically and experimentally suggested phenomenon 
indicating the less rescatterings of $\phi$-mesons, 
violation of mass ordering in $v_2(p_T)$,
has been interpreted as a results of the effects on mean $p_T$ and $v_2$.
These results at the RHIC energy had been discussed in Ref.~\cite{Takeuchi:2015ana}.
Now we have shown that this behaviour also appears at the LHC energy.
By considering these results,
we claim that the multi-strange hadrons can be utilised as ``penetrating" probes of the QGP 
in high-energy heavy-ion collisions.

\section*{Acknowledgement}
This work was supported by JSPS KAKENHI Grant Numbers 
12J08554 (K.M.) and 25400269 (T.H.), and by BMBF under contract no. 06FY9092 (P.H.).



%% References


\bibliographystyle{elsarticle-num}
\bibliography{<your-bib-database>}



\end{document}